\documentstyle[prb,aps,twocolumn]{revtex}
\large
\begin{document}
\draft
\twocolumn[\hsize\textwidth\columnwidth\hsize\csname
@twocolumnfalse\endcsname

\title{Separation of the magnetic phases at the N\'{e}el point
in the diluted spin-Peierls magnet CuGeO$_3$}

\author{V.~N.~Glazkov, A.~I.~Smirnov}
\address{P.~L.~Kapitza  Institute for  Physical  Problems  RAS,
117334 Moscow, Russia}

\author{K.~Uchinokura, T.~Masuda}
\address{
Department of Advanced Materials Science, The University of
        Tokyo, 7-3-1 Hongo, Bunkyo-ku, Tokyo 113-8656,  Japan
}

\date{\today}
\maketitle

\begin{abstract}
\widetext
\leftskip 54.8pt
\rightskip 54.8pt
The impurity induced antiferromagnetic ordering of the doped
spin-Peierls magnet Cu$_{1-x}$Mg$_x$GeO$_3$ was studied by
electron spin resonance (ESR) technique. Crystals with the Mg
concentration $x <$ 4\% demonstrate
a coexistence of
paramagnetic and antiferromagnetic ESR modes.  This coexistence
indicates the separation of a macroscopically uniform sample
in the paramagnetic and antiferromagnetic phases.  In the presence of
the long-range spin-Peierls order (in a sample with  $x=1.71$\%) the
volume of the antiferromagnetic phase immediately below the N\'{e}el
point $T_N$ is much smaller than the volume of the paramagnetic phase.
In the presence of the short-range spin-Peierls order (in samples with
$x=2.88$\%,  $x= 3.2$\%) there are comparable volumes of
paramagnetic and antiferromagnetic phases  at $T=T_N$.
The fraction of the antiferromagnetic phase
increases with lowering temperature.  In the absence of the
spin-Peierls dimerization (at $x=4.57$\%) the whole sample
exhibits the transition into the antiferromagnetic state and there is
no phase separation.  The phase separation is explained by
the consideration of clusters of staggered magnetization located near
impurity atoms.  In this model the areas occupied by coherently
correlated spins expand with decreasing temperature and the percolation
of the ordered area through a macroscopic distance occurs.

\end{abstract}

\pacs{PACS numbers:  75.10.Jm, 76.50.+g, 75.50.Ee}

]

\narrowtext

\section{Introduction}

Quasi-one-dimensional magnet CuGeO$_3$ is a unique inorganic
compound demonstrating a spin-Peierls phase
transition.\cite{Hase}
The  spin-Peierls transition may occur in a crystal
containing spin $S=1/2$ antiferromagnetic chains due to the
spin-lattice instability with respect to the dimerization of magnetic
ions.\cite{Pytte}  Below the transition temperature
$T_{SP}$=14.5~K the lattice period along the chain direction
becomes doubled and the exchange integral alternates taking in
turn two values $J \pm \delta J$. Due to this alternation the
ground state is a singlet separated from the excited triplet
states by an energy gap $\Delta $=2~meV.\cite{Nishi} Thus at
low temperatures pure crystals appear to be almost nonmagnetic
and a small residual magnetic susceptibility is provided only by
defects.  The amplitude of the atomic displacements
resulting in the dimerization can be chosen as the order
parameter of the spin-Peierls phase.  Impurities substituting
magnetic or nonmagnetic ions disturb the homogeneity of the
spin-Peierls phase in CuGeO$_3$.  The doping diminishes the
transition temperature and results in an
antiferromagnetic long-range
ordering.\cite{Hase3,Hase4,Oseroff,Regnault,Lussier}  The
spin-Peierls dimerization and the impurity-induced magnetic
order were found to coexist at low impurity concentration $x$.
Stimulation of the long-range antiferromagnetic order by
impurities was explained in
Refs.~\CITE{Fukuyama,Khomskii,Khomskii2}.  The violation of the
dimerization around  an impurity results in the formation of a
soliton-like spin cluster with an antiferromagnetic correlation
of neighboring spins and staggered magnetization.  Overlapping of
clusters and the weak interchain exchange result in the long-range
three-dimensional antiferromagnetic order.

The phase transition to the antiferromagnetic state and  $T-x$
phase diagram were studied for different types of doping
atoms.
\cite{Renard,Hase5,Coad1,Martin,Grenier,Manabe,Masuda,Nakao}
For antiferromagnetic phases stimulated by the impurities of
Zn, Mg, Si the easy axis
 of the ordered state is $c$ axis, and
for Ni-doping - the $a$ axis. The phase diagram  contains
areas of a uniform (i.e.,  without dimerization) paramagnet, of a
dimerized paramagnet (i.e., of spin-Peierls state), and of the
antiferromagnetic state. It was shown experimentally that there
is no threshold concentration of impurities for the occurrence
of the magnetic order.\cite{Manabe} Further detailed
investigations revealed subtle features of the phase diagram  shown
on the Fig.~1 for  Mg-doped crystals of CuGeO$_3$ following
Ref.~\CITE{Masuda2}.  The first order phase transition between
dimerized and uniform antiferromagnetic phases at
a critical value of concentration was found.\cite{Masuda,Masuda2}
The transformation from the dimerized to the uniform  phase in Mg-
doped samples occurs in a concentration range between 2.37\% and
2.71\%. The uniform phase demonstrates a higher value of the N\'{e}el
temperature.
Further, the state with a
short-range spin-Peierls order was detected for Mg-doped samples and
the transition between short range and long-range ordered spin-Peierls
phases was found.\cite{Masuda2,Wang} Finally, the reentrant phase
transition into the undimerized state  from the dimerized
antiferromagnetic phase  was observed.\cite{Kiryukhin} A
theoretical consideration  of the transition between
dimerized and uniform antiferromagnetic phases and of the reentrant
transition was given in Ref.~\CITE{Khomskii2}.

  This variety of phases is caused by the competition between
the gapped dimerized state and the antiferromagnetic state which
is gapless in the exchange approximation.  The spin-Peierls
state does not  allow the three dimensional antiferromagnetic
ordering in the pure compound and, on the other  hand,
impurities  restore the antiferromagnetic correlations and
suppress the spin-Peierls dimerization.

The goal of this work is an ESR study of magnetic properties of
different antiferromagnetic phases and of the phase transitions
at various phase boundaries of the phase diagram.  The previous
ESR investigations have revealed the multispin nature of
clusters formed near impurity ions,\cite{Glazkov} and the
gap in the zero-field ESR frequency in the antiferromagnetic
phase.\cite{Glazkov,Hase2,Smirnov,Nojiri} Thus the
evolution from isolated clusters with local staggered
magnetization  to the long-range
antiferromagnetic order may be followed using the ESR technique.  We
used single crystal samples of Cu$_{1-x}$Mg$_x$GeO$_3$ from the
measurements \cite{Masuda,Masuda2,Kiryukhin} of the phase diagram shown
on Fig.~1, or the samples grown by the same method on the same
installation.  The phase transitions studied in the present work
are marked on Fig.~1.

As a result of this study we found that the formation of the
antiferromagnetic order at small impurity concentrations is
accompanied by the microscopic phase separation into the
paramagnetic and antiferromagnetic phases and this phase
separation differs in temperature evolution for
antiferromagnetic phases coexisting with a long-range and
a short-range ordered spin-Peierls background.

\section{Samples and experimental details}

The  single  crystals  of  the Cu$_{1-x}$Mg$_x$GeO$_3$
with  $x=$~0, 1.71\%, 2.88\%, 3.20\%, 4.57\% were  grown  by
the  floating  zone method. The impurity  distribution
was  checked by inductively coupled  plasma  atomic
emission  spectroscopy technique and was found  to  be
uniform  within  0.1\%   (see Ref.~\CITE{Masuda2}).  We
used  single  crystals with the  dimensions  of  about
1$\times$2$\times$2 mm.

The concentration of the residual magnetic defects (both of the
structure and impurity type) may be estimated from measurements
of the ESR intensity of a  nominally pure ($x=0$) sample (see the inset
of Fig.~7).  This intensity rapidly decreases below the transition
temperature due to the freezing out of the gapped triplet
excitations. The minimum values of the ESR intensity and of the
static susceptibility observed at 4.5~K  equal 5\% of these
values  at $T_{SP}$. This intensity and susceptibility correspond
to the concentration of  the residual defects $x_{def}\sim 0.07$\% per
Cu-ion.

 The ESR spectra were taken by means of a spectrometer with a
set of transmission-type  resonators.  Measurements   were
carried out in the  frequency  range  9~--~75~GHz  at
temperatures 1.5~--~15~K.  The magnetic resonance absorption line
was recorded as  a dependence of the transmitted microwave power
on  the applied magnetic   field.  The  reduction   of   the
transmitted signal  is proportional to the
microwave power absorbed  by the sample.  For  the paramagnetic
state  the integrated  intensity  of absorption is proportional
to the static susceptibility.

\section{Experimental results}

The  temperature  evolution of the ESR  line  for  the sample
with the impurity concentration $x$~=~4.57\%  is typical of an
antiferromagnet: at the decrease of the temperature starting
from the N\'{e}el point ($T=T_N$) the single resonance line shifts
to lower fields when the magnetic field is perpendicular to the
easy axis  of spin ordering (see Fig.~2).

The  ESR  lines  of  the samples with  $x=1.71$\%  and $x=2.88$\%  are
shown  on the  Fig.~3  and  Fig.~4.  At $T=T_N$ the resonance line
splits into  two  spectral components.  Similar splitting was  observed
for  the sample with $x=3.20$\%. Below the N\'{e}el temperature the
ESR  line is well described as a sum of  the  two Lorentzian
components (see  inset  of  Fig.~3).   The component  shifting  to
lower fields  with  decreasing temperature has a nonlinear
field-dependence of the resonance frequency with strong anisotropy,
shown in Fig.~5  for the sample with $x=3.20$\%.  We note this spectral
component as an antiferromagnetic resonance line since this
frequency-field dependence with two gaps  is typical for two-axes
antiferromagnets.\cite{AFMR}  The other spectral component has a linear
frequency-field dependence with a temperature independent $g$-factor.
We note this absorption mode as a paramagnetic resonance  corresponding
to the $g$-factor values $g_a=2.14$ and $g_b=2.21$ measured for the
field orientations along $a$ and $b$ axes correspondingly.  These
values coincide within the experimental errors with $g$-factors
obtained for pure compound.

The temperature dependence of the resonance fields at a fixed
frequency is shown in Fig.~6 for the sample with $x=2.88$\%.
The temperature when the low-field
line starts to shift from the paramagnetic  resonance position
corresponds well  to  the value of $T_N$ obtained by
susceptibility measurements.\cite{Nakao,Masuda2}

To obtain the ESR spectra with two resolvable components one
should take the microwave frequency close to the
antiferromagnetic resonance gap.  In this case the difference
between the paramagnetic resonance field and that of a gapped
antiferromagnetic resonance mode will be more significant,
helping to resolve two spectral components.  As shown on
Fig.~5 the spectrum of antiferromagnetic resonance of the
doped CuGeO$_3$ has two branches with different gaps.  The
observation of the first or the second branch by the field-sweep
technique  depends on the field orientation.  Therefore to meet
the condition mentioned we selected not only the microwave
frequency from the set of the resonant frequencies of the
microwave resonator, but also the orientation of the magnetic
field with respect to crystal axes.  Because the values of gaps
depend also on Mg-concentration, the data for different samples
are obtained at different microwave frequencies and
orientations of the external field.

The   temperature   dependences  of   the   integrated
intensities of both  components are shown on  Fig.~7 for
$x=1.71$\%, and on Fig.~8  for $x=2.88$\% and 3.20\%.  The
remarkable feature of the two-component ESR spectrum is the
large intensity of the paramagnetic line  in a temperature
range below the
N\'{e}el point.  For the low concentration $x=1.71$\% the
intensity of the paramagnetic line  below the N\'{e}el point is
close to the integral intensity above $T_N$, and the intensity of
the antiferromagnetic resonance mode is much smaller than that
of the paramagnetic mode.  For concentrations  of $x=2.88$\% and
3.20\%  on the contrary,
the intensity of the
antiferromagnetic resonance is larger than the intensity of
paramagnetic mode.  For the concentration $x=4.57$\% there is no
distinguishable paramagnetic mode below N\'{e}el temperature.  We
ascribe the whole intensity to the antiferromagnetic mode and take the
intensity of the paramagnetic resonance as zero.  The $x-$dependences
of the relative intensities of the paramagnetic and antiferromagnetic
modes extrapolated to $T_N$ from low temperatures are plotted in Fig.~9.
Because of  significant errors occurring near $T_N$ in the
determination of the intensity of a weak and wide antiferromagnetic
component near the narrow paramagnetic line the intensity of the
antiferromagnetic component tends to  be overestimated.

\section{Discussion}

\subsection{Possible reasons for the paramagnetic mode in the
antiferromagnetic state}

The identical values of the $g$-factor of the paramagnetic
spectral component below and above $T_N$ indicate that this ESR
signal is due to Cu$^{2+}$ ions.  The intensity of this component
decreases with decreasing temperature, therefore it can not be
ascribed to isolated Cu ions located at the  surface or at structure
defects. The isolated spins would show an increase in
intensity according to the Curie-law.

The paramagnetic resonance signal might be  due to the triplet
excitations of the spin-Peierls magnet which are present both in pure
and doped crystals. \cite{Regnault,Martin,Sasago}
However, due to the gap in the triplet excitations spectrum the
intensity of this signal should rapidly drop with cooling below
$T_{SP}$.
To estimate the ESR intensity provided by the triplet excitations we
take the whole intensity in the temperature range  $T_N<T<T_{SP}$ to be
due to the additive contributions of the impurity Curie-like part and
the triplet excitation part:  \cite{Bulaevski}

\begin{equation} I(T) =
\frac{C}{T-\theta}+\frac{D}{T} \exp (-\frac{\Delta}{kT})
\label{IvsT}
\end{equation}

Here $C,\theta  $ are the Curie constant and Curie-Weiss temperature,
$D$ is a constant depending on dimerization.  The results of the
fitting of the ESR intensity by this formula for the pure sample and
for the doped sample with $x=1.71$\% in the temperature range between
$T_N$ and $T_{SP}$ are shown on  the inset of Fig.~7.  We find that for
$x=1.71$\% the triplet part of the ESR intensity at the N\'{e}el point
should be of about of 10$^{-3}$ of the observed value and thus could
not provide the paramagnetic resonance signal observed.

 For larger concentrations the triplet part of the intensity and
susceptibility is not well pronounced (there is no evident drop of the
magnetic susceptibility \cite{Masuda} at lowering temperature).
Nevertheless we can exclude the triplet excitations for the following
reasons.  At $x=3$\% the triplet excitations are
overdamped,\cite{Martin,Sasago} hence they should  strongly interact
with the spins near impurities and the coupling of the
antiferromagnetic resonance and of the triplet mode  should occur.
Depending on the coupling parameter, the resulting ESR spectrum of two
exchange coupled spin systems (see, e.g.  Ref.~\CITE{Anderson}) should
consist of a single line at the intermediate frequency or of two
separate lines with frequencies varying simultaneously at the change of
external parameters or of the coupling coefficient. Examples of the ESR
spectra of the exchange coupled systems including a spin-Peierls
magnet, an antiferromagnet with impurities and thermally activated
spins in organic molecules may be found in
Refs.~\CITE{Glazkov,Glazkov2,Honda,Smirnov2,Dumesh,Chesnut}.  In
the present study we observe two-component spectrum with  constant
field of the paramagnetic mode and with the  shift of the
antiferromagnetic line. Consequently  there are no traces of the mutual
influence  of two modes which should occur if the paramagnetic mode
would be due to the triplet excitations.  Thus for $x>2$\% the
contribution of triplet excitations to ESR intensity  at the N\'{e}el
temperature is also negligible.

Further we prove that the observed two-component ESR
signals couldn't be ascribed to a trivial inhomogeneous distribution of
the Mg-concentration, and, hence to different values of $T_N$ in
different parts  of the sample.  The distribution of the N\'{e}el
temperature would result  in the wide   band  of absorption, while
the observed antiferromagnetic resonance absorption is well described
by a single Lorentzian. Besides that the  range of the distribution of
$x$  which is necessary to account  for the paramagnetic phase is much
wider than obtained  in control measurements.  For example,  to  have
the N\'{e}el points in the range 1.5$-$2.25 K  where both signals are
present for the $x=$1.71\% sample,  we should imagine the concentration
distribution  in the range  1.3$-$2.2\%.   The width  of this
range  is  much larger  than   0.1\%  obtained in concentration
measurements.  The    well     defined singularities   on
temperature dependences   of   the susceptibility
\cite{Masuda2}  and of the resonance field (see Fig.~6)  prove
that the samples are macroscopically uniform.  The width of the
transition  to the N\'{e}el   state may  be   estimated from
the susceptibility and ESR data, and we see that it is not larger than
0.1~K.

Summarizing the above analysis of possible reasons for the paramagnetic
resonance mode below the N\'{e}el point we state that the
observed paramagnetic resonance signal can not be ascribed to isolated
uncontrolled magnetic defects, to triplet excitations or to an
inhomogeneous distribution of impurities.

The  uniform antiferromagnet is characterized  by a
single order parameter and should exhibit only antiferromagnetic
resonance modes corresponding to a certain type of ordering.  The
experiments on numerous antiferromagnets show that at the N\'{e}el
point the paramagnetic resonance converts in the antiferromagnetic
resonance, if the frequency is larger than the gap of the
antiferromagnetic resonance, as we observe for the sample with
$x=$4.57\%.  If the frequency is  much smaller than the gap value, the
ESR signal in a usual antiferromagnet disappears at the N\'{e}el point.
In both cases there is no ESR signal on the paramagnetic resonance
frequency below $T_N$.  For the existence of an additional paramagnetic
resonance mode in the uniform antiferromagnet an additional spin degree
of freedom is necessary.  A hypothetical additional mode related to
Cu$^{2+}$ spins in our case  should be coupled with the
antiferromagnetic resonance mode. As described above, we excluded the
spin degrees of freedom of the isolated defects and of the triplet
excitations, and we detected the absence of any coupling of two modes.
Therefore we come to the conclusion that our system is not
microscopically uniform and the coexisting paramagnetic and
antiferromagnetic resonance signals should originate from different
areas of the sample. Thus a microscopic separation  in the paramagnetic
and antiferromagnetic phases takes place in macroscopically uniform
samples with low impurity concentrations.

\subsection{Geometrical model}

We  explain  the  microscopic
phase  separation at low impurity concentrations,  when long-range
dimerization order occurs, considering the regions of antiferromagnetically
correlated spins \cite{Fukuyama,Khomskii} (spin clusters) appearing
near all impurity atoms. The spins within these spin clusters have
nonzero average spin projections in form of staggered magnetization,
thus the local N\'{e}el order parameter can be introduced. Besides that
a cluster has a net magnetic moment equal to $\mu_B$. The correlation
length along the spin chain $\xi_c$ is estimated to be of about 10
interspin distances.\cite{Khomskii,Kojima} The interchain exchange
integrals should result in the spin-spin correlations transverse to the
chain direction.  The transverse correlation lengths $\xi_{a,b}$ may be
estimated analogous to the estimation of the longitudional correlation
length \cite{corrlength,Regnault2}  as follows: $\xi_{i} \sim
v_{i}/\Delta$, here $v_{i} \sim J_{i}l_{i}$ is the spinons velocity
with $J_{i}$ being the interchain exchange integrals along the
directions   $a$  and $b$ , $l_i$ are  the lattice constants  along
these  transverse directions.  Thus the cluster at $T=0$
may be considered  as a 3D region with the staggered magnetization
located near the impurity and with the exponential decay of this
staggered magnetization at moving away from the impurity.
At finite temperature
the coherence of the antiferromagnetic order parameter which is
spatially variable on the wing of the cluster  will be destroyed by
thermal fluctuations. The distance $L$ in the chain direction of the
region of the coherent antiferromagnetic order parameter may be
estimated from the relation:

\begin{equation} \label{size}
k_BT=JS^2\exp\{-2L/\xi_c\}.
\end{equation}

The distances of the coherence along transverse directions are
taken  to be equal to $\frac{\xi_i}{\xi_c}L$.
Therefore  at  finite temperature the area of the spatially
coherent antiferromagnetic order may be considered as an
antiferromagnetic drop  of the ellipsoidal form with fixed
boundaries.  This drop is elongated along the spin
chain direction and the ratio of the drop dimensions along and
transverse to the chain is of about the ratio of  corresponding
exchange integrals.  For CuGeO$_3$ we have this ratio according to
Refs.~\CITE{Nishi,Regnault2}:  $J_b/J$=0.11, $J_a/J$=-0.011.

The size of an ellipsoidal drop
enlarges when temperature is lowering according to the relation
(\ref{size}). A drop has a net magnetic moment of $\mu_B$.  The drops
are placed in space at random with the density of drops corresponding
to the value of the concentration of impurities. At high temperatures
when the  drops are small and do not overlap, the antiferromagnetic order
parameter is nonzero within ellipsoids and zero outside them.  The
phases of the local order parameters of different drops are not
coherent and the model shows no long-range antiferromagnetic
order.  At lowering temperature the  drops grow and
begin to overlap.  The order parameter in the
overlapped drops (a conglomerate of drops) is coherent, thus
large areas with coherent antiferromagnetic ordering appear
due to the formation of conglomerates.
This model is illustrated in Fig.~10 by drawing the expanding drops in
two dimensional space.  One can see on Fig.~10 how the area of the
coherent antiferromagnetic order penetrates through a macroscopic
distance along both coordinates.  We see that in
the process of the formation of large ordered regions there are also
islands of zero order parameter and ordered drops within these islands.
The order parameter propagates in the process of percolation of the
ordered phase through a macroscopic distance.  The percolation  occurs
here as it takes place in the phenomenon of the percolation through the
randomly placed interpenetrating spheres at a critical value of the
volume fraction occupied by spheres.\cite{ShanteKirkpatrick}

\subsection{Correspondence between the geometric model and experimental
results}

The observed results for $x=1.71\%$ sample may be explained
on the basis of the described model as follows.  At a   temperature
above the N\'{e}el point the drops are  small  and  hence isolated
(Fig.~10a).
All  drops contribute to  the
Curie-type susceptibility due to their magnetic moments  equal
to $\mu_B$. The dimerized  spin-Peierls matrix  (white  area on
Fig.~10) also  contributes to  the magnetic susceptibility via triplet
excitations, but this contribution exponentially  decreases with
decreasing temperature  and becomes negligible near the N\'{e}el point.
The isolated drops contribute to the intensity of the paramagnetic
resonance mode because of the Zeeman splitting of the energy levels due
to the net magnetic moment of the drop $\mu_B$, therefore at $T \geq
T_N$ we should observe a single ESR line with $g$-factor close to that
of Cu ions.

At  the phase transition at the percolation point
there is a phase separation: there are  intersecting  threads
of antiferromagnetically  ordered
phase  (marked with black color on Fig.~10c), dimerized spin-Peierls
matrix (white) and still remaining isolated drops (gray).  We should
observe both the antiferromagnetic resonance from the threads of
ordered phase and the paramagnetic resonance signal from the drops
isolated within the dimerized matrix.

As the temperature  decreases further,  the
isolated drops join the antiferromagnetic phase
and the  volume of ordered phase increases.
The  volume  of  the
paramagnetic  phase  reduces and the  intensity  of  the
paramagnetic   signal should   diminish  with   temperature
decrease.

Magnetic  measurements (both static and ESR) should detect  the
transition into the antiferromagnetically ordered  state
when  antiferromagnetic susceptibility which is  due  to
the staggered magnetization within the conglomerates of drops
prevails over   the  paramagnetic  susceptibility of this
conglomerates.  We may
estimate the number $N$ of spins in a conglomerate  at this moment.
The paramagnetic susceptibility given by the Curie-law is of the order
of $\mu_B^2/(k_BT)$ for a conglomerate.  Antiferromagnetic
susceptibility $\chi_{\perp}$ per spin is of the order of $\mu_B^2/J$.
Thus we obtain

\begin{equation} N\sim\frac{J}{k_BT} \sim 100
\label{number}
\end{equation}

The   scenario  described  above  explains  mainly   the observed phase
separation below the N\'{e}el temperature in the impurity stimulated
antiferromagnetic phase of the spin-Peierls  magnet in the low
concentration range. The dimerized background and the random
distribution of impurities are of importance for this scenario.

The ratio of the intensity of paramagnetic component just below
the N\'{e}el temperature to the intensity of the ESR line above
the N\'{e}el  temperature  is a measure  of  the  sample volume
occupied by the paramagnetic phase.  The analogous ratio of the
antiferromagnetic component represents the fraction of the
antiferromagnetic phase.
The results presented on Fig.~9 show that the
antiferromagnetic fraction  just below $T_N$ is small
(between zero and 25\% of the
sample volume) for the sample with $x=1.71$\%. For the samples with
$x=2.88$ and 3.2\% this volume is larger and exceeds a half of the
sample volume, and for $x=4.57$ \% the whole sample becomes
antiferromagnetic at the N\'{e}el point.  Taking into account that
these concentrations lie in different parts of the phase diagram
(Fig.~1) we conclude from our results:
i) the volume of the antiferromagnetic phase at $T=T_N$ is small
when the ordering takes place at the long-range
spin-Peierls order; ii) at the short-range spin-Peierls order there are
comparable volumes of two phases at the  $T=T_N$; iii) in the absence
of spin-Peierls dimerization the whole sample becomes ordered at the
transition temperature. The first conclusion is in agreement with
geometrical model: the volume of a percolating thread is smaller
than the sample volume (see Fig.~10c). This observation is to be
compared with the result of the percolation theory for
interpenetrating spheres placed at random, giving
the volume fraction at the percolation point $v_c$=0.286 for three
dimensions and 0.675 for two dimensions.

We can not extrapolate the constructed model to the situation with
short range dimerization order. Nevertheless it is natural to propose
that in the corresponding concentration range, the behavior of the
system should be intermediate between the behaviors of dimerized and
uniform crystals with defects. This proposition  is shown  in Fig.~9
with straight lines interpolating the fraction of the antiferromagnetic
phase from a small value  at long range spin-Peierls order to unity at
undimerized phase.  This hypothesis is in a qualitative agreement with
our observations. Further detailed investigations of the amount of the
ordered phase just below the N\'{e}el point for different concentration
are of great interest.

It is worth to note that in several previous investigations  of the
antiferromagnetic phase in doped CuGeO$_3$ the two-component signal was
not observed.\cite{Glazkov,Hase2,Nojiri} The single line
may be explained here either by frequencies far from the
antiferromagnetic resonance gaps,\cite{Nojiri} or by large
impurity concentrations which suppress the
dimerization.\cite{Glazkov,Hase2}
The two-component line with the spectra and
temperature dependences  analogues to the reported in the present work
was observed in experiments with
Cu$_{0.98}$Zn$_{0.02}$GeO$_3$.\cite{Smirnov}

\section{Conclusion}

ESR  measurements  reveal  the  microscopic   phase
separation  at  the  impurity  induced  antiferromagnetic
ordering  in  the  spin-Peierls  magnet  CuGeO$_3$.   The
temperature  evolution of the ordered phase  volume  with
the small volume fraction  at the N\'{e}el point indicates
the  percolating character of the antiferromagnetic phase
transition    at    low   doping    level    when     the
antiferromagnetic and spin-Peierls order coexist.

\section{Acknowledgements}

The  authors  are  indebted  to  S.~S.~Sosin  for  valuable
discussions.

The work was supported
 by the joint grant of the Russian Foundation for Basic Research and
 Deutsche Forschung Gesellschaft (project No. 01-02-04007), by
 INTAS project No. 99-0155 and by Award No. RP1-2097 of the U.S.
 Civilian Research and Development Foundation for the Independent
 States of the former Soviet union (CRDF).  One of the authors (V.G.)
 thanks Forshungscentrum Julich Gmbh for continuous support of his
 work.

\section{Figure captions}
\begin{itemize}

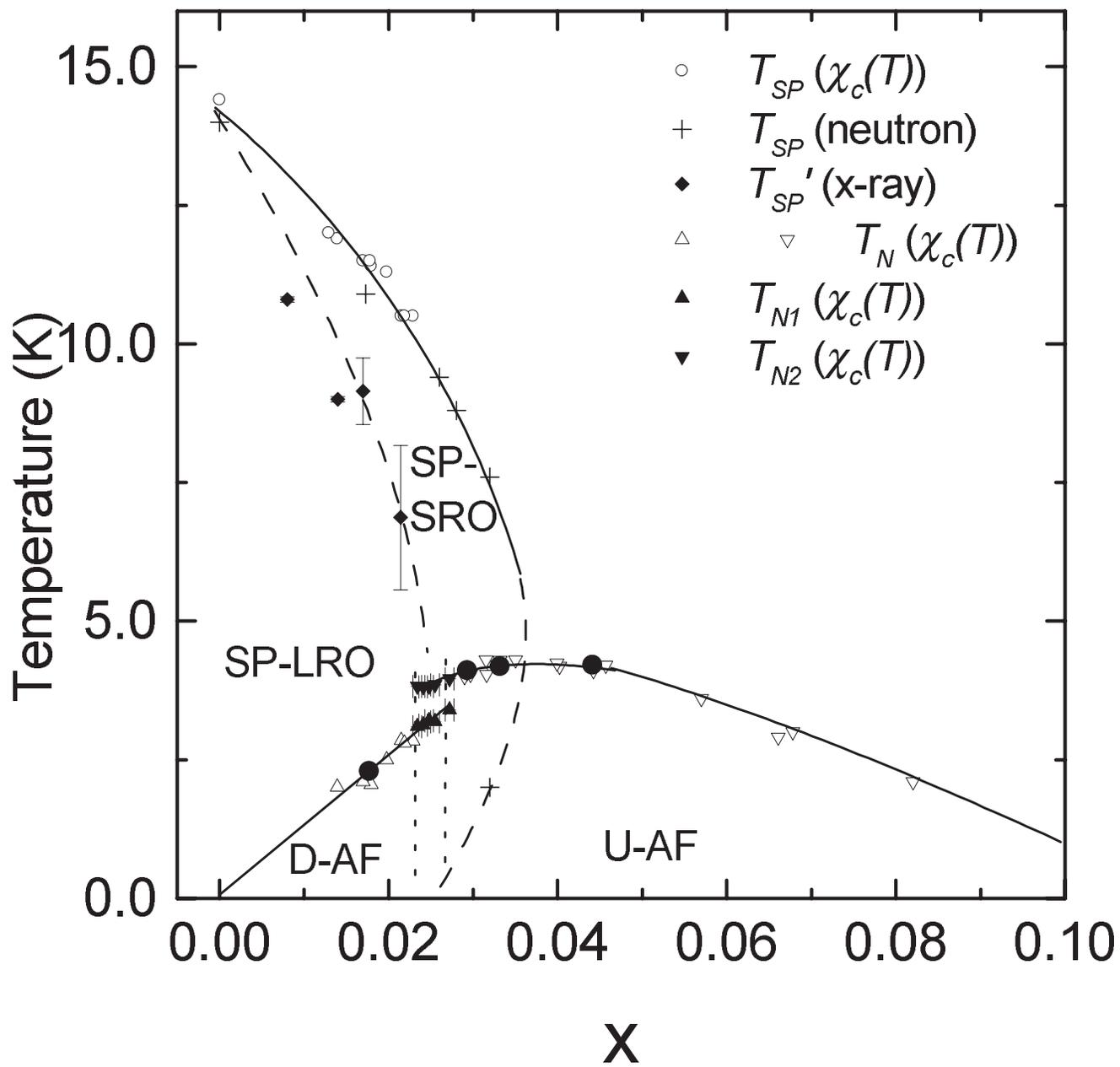
\item[Fig.1]   Phase  diagram  of  the  diluted   Cu$_{1-
x}$Mg$_x$GeO$_3$ following Ref. \CITE{Masuda2}. The phase
transitions  studied in the present work are  marked  by
signs $\bullet$.

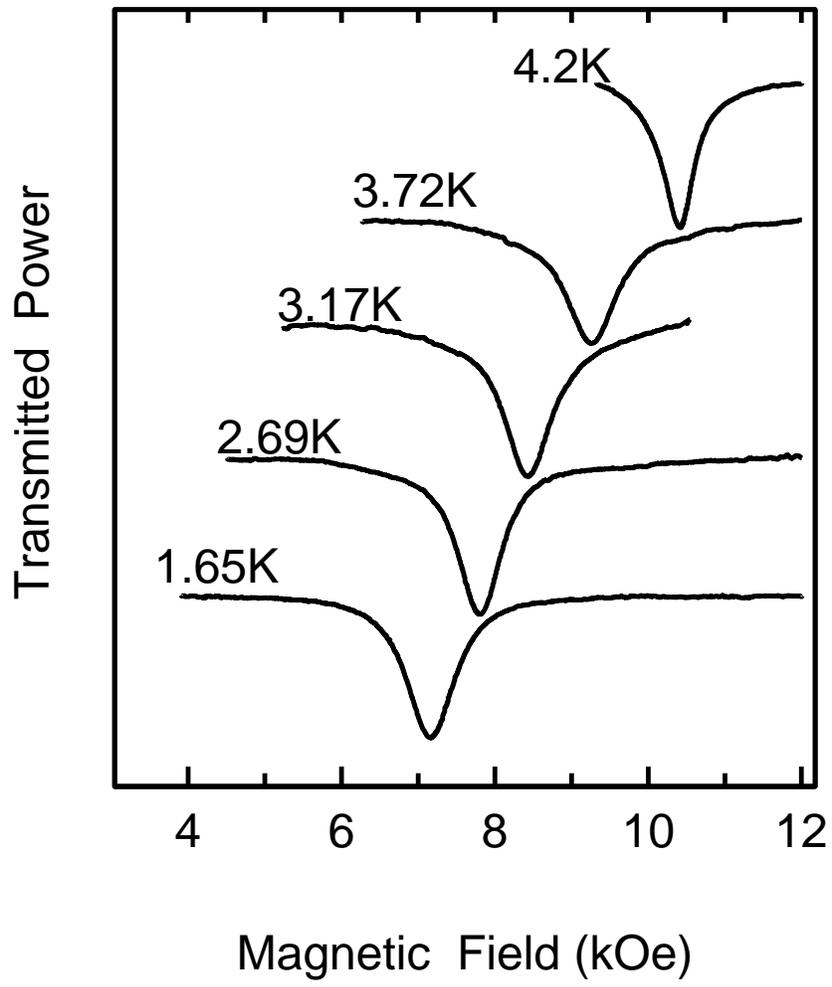
\item[Fig.2] Evolution of the ESR
line for the sample containing 4.57\% Mg.
${\bf H} \parallel a$, $f=31$~GHz, $T_N$=4.20K.

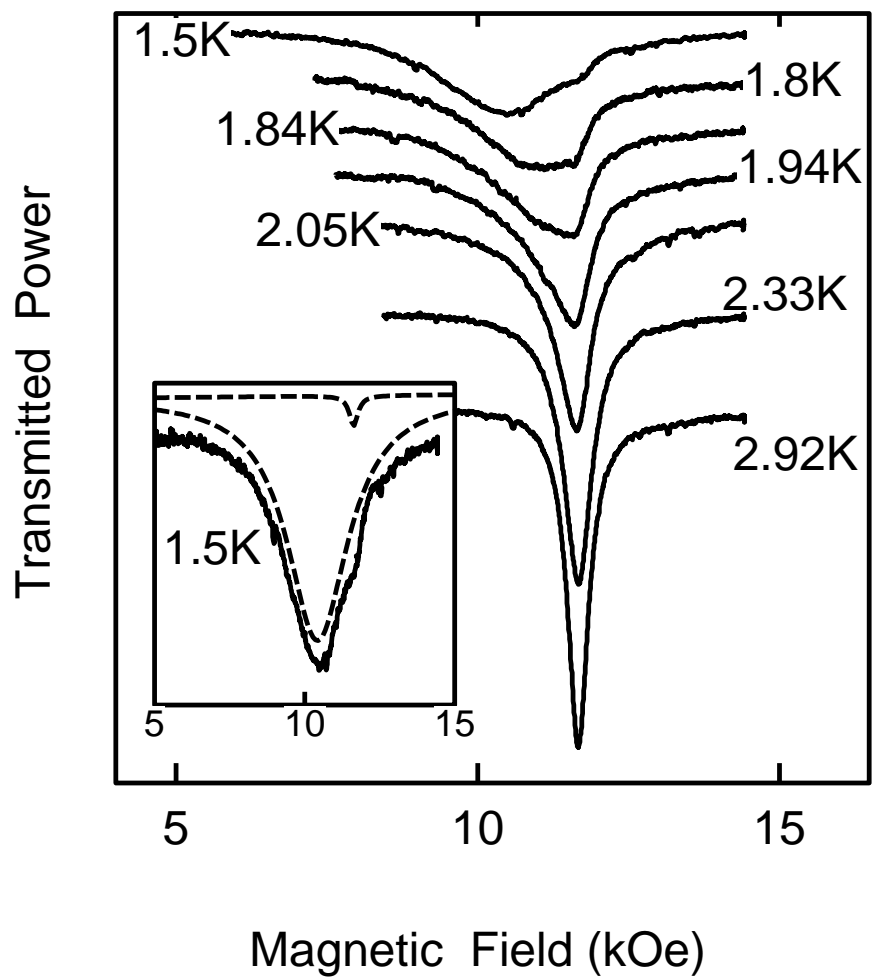
\item[Fig.3] Evolution of the ESR
line for the sample containing 1.71\% Mg.
$  {\bf H} \parallel b$, $f=36$~GHz, $T_N$=2.25K. Inset: ESR
line at 1.5K and the Lorentzian components.

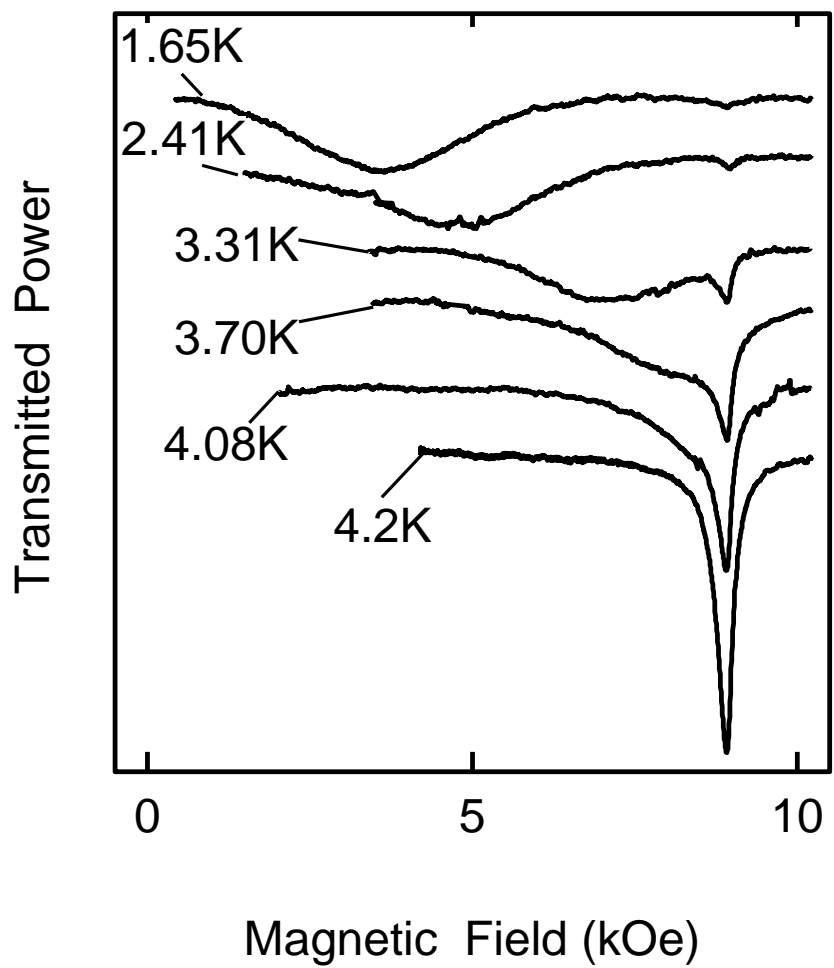
\item[Fig.4] Evolution of the ESR line
for  the  sample  containing  2.88\%  Mg. $  {\bf
H}\parallel a$, $f=26.3$~GHz, T$_N$=4.14K.

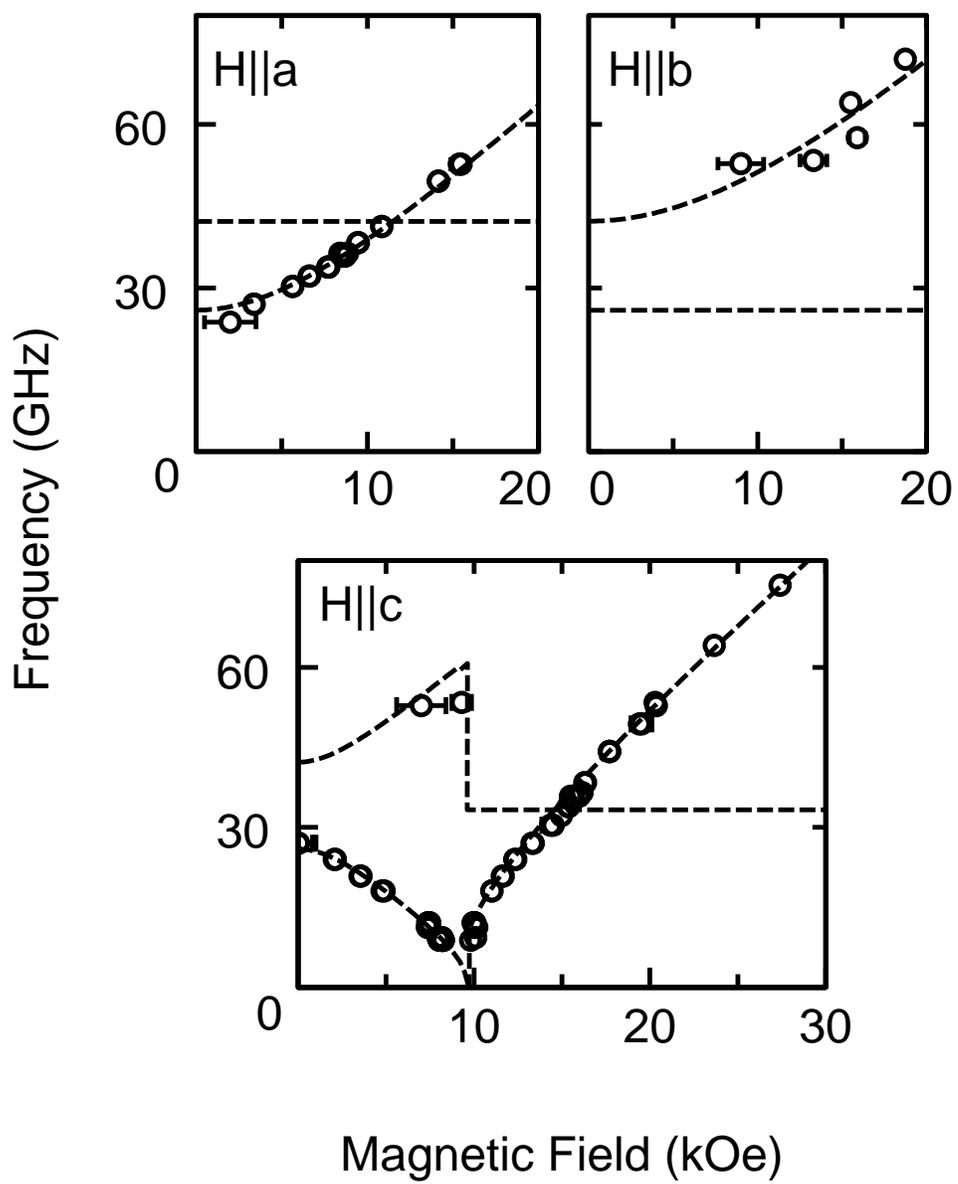
\item[Fig.5] The spectrum of the antiferromagnetic resonance of the
3.2\% Mg doped sample at $T=1.8$~K for three principal directions of
the magnetic field with respect to crystal axes. Dashed lines
represent the theoretical calculations following Ref. \CITE{AFMR}.

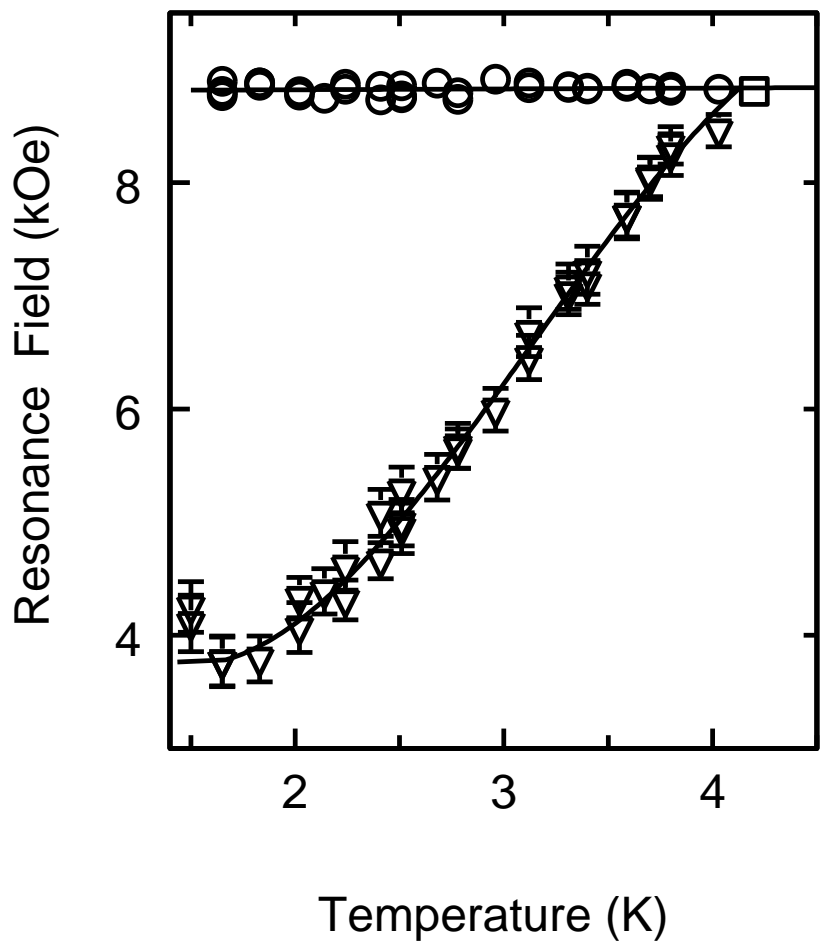
\item[Fig.6] The temperature dependences of
the 26.34~GHz ESR fields for the sample containing 2.88\%
Mg at ${\bf H} \parallel a$.
The signs $\bigtriangledown$ correspond to the
antiferromagnetic  resonance  and  $\bigcirc$  to   the
paramagnetic resonance, $\Box$ - magnetic resonance above $T_N$.

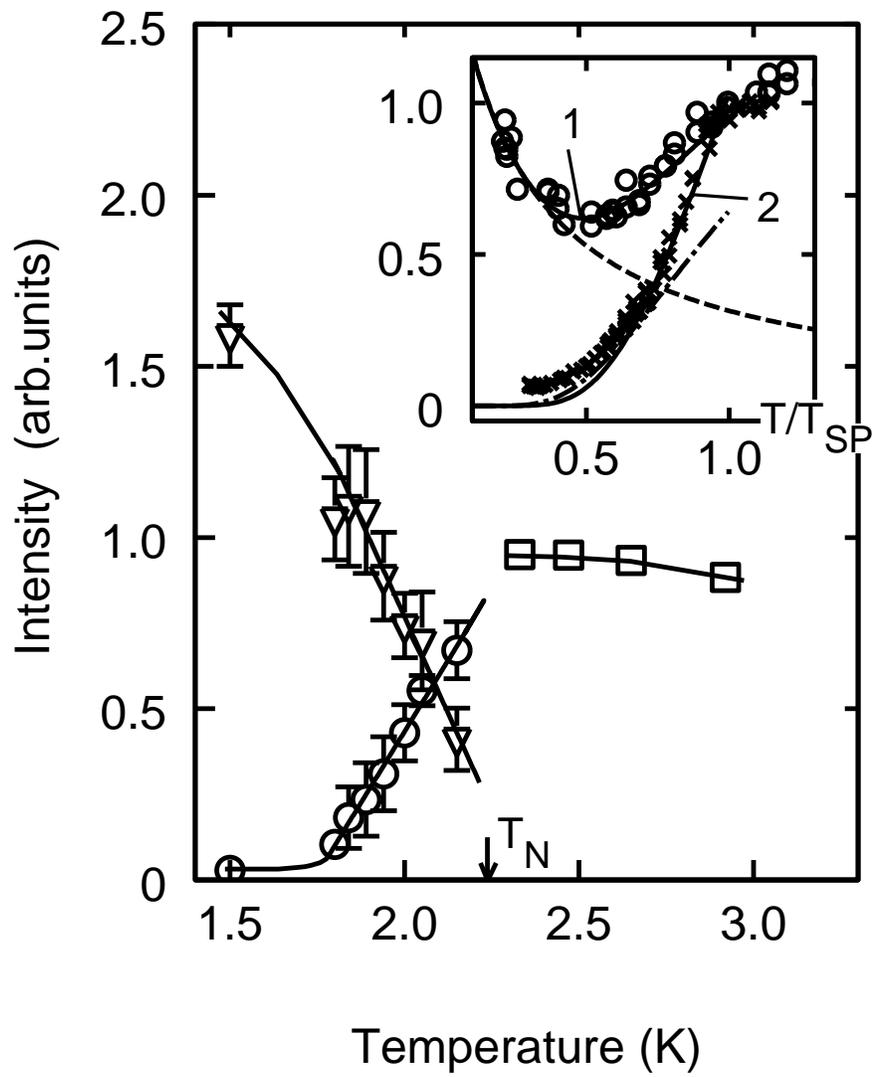
\item[Fig.7] Temperature dependences of the intensities
of 36GHz ESR spectral  components at ${\bf H} \parallel b$.  for the
sample  with  impurity concentration  $x$=1.71\%.  The  signs:
$\Box$-above $T_N$,    $\bigtriangledown$    correspond    to    the
antiferromagnetic  resonance  and  $\bigcirc$  to   the
paramagnetic  resonance below $T_N$.  Solid  lines  are
guide-to-eyes.
Inset: 36GHz ESR intensities at ${\bf H} \parallel b$ for  pure
($\times$)  and  1.71\% Mg-doped
($\bigcirc$)  samples
at ${\bf H} \parallel b$.
Solid line 1 represents the fit by formula (\ref{IvsT})
for $x=1.71$\%, dashed and dotted lines are the first and
second terms in (\ref{IvsT}), solid line 2 is the second term
obtained from the fitting by formula (\ref{IvsT}) for the pure sample.

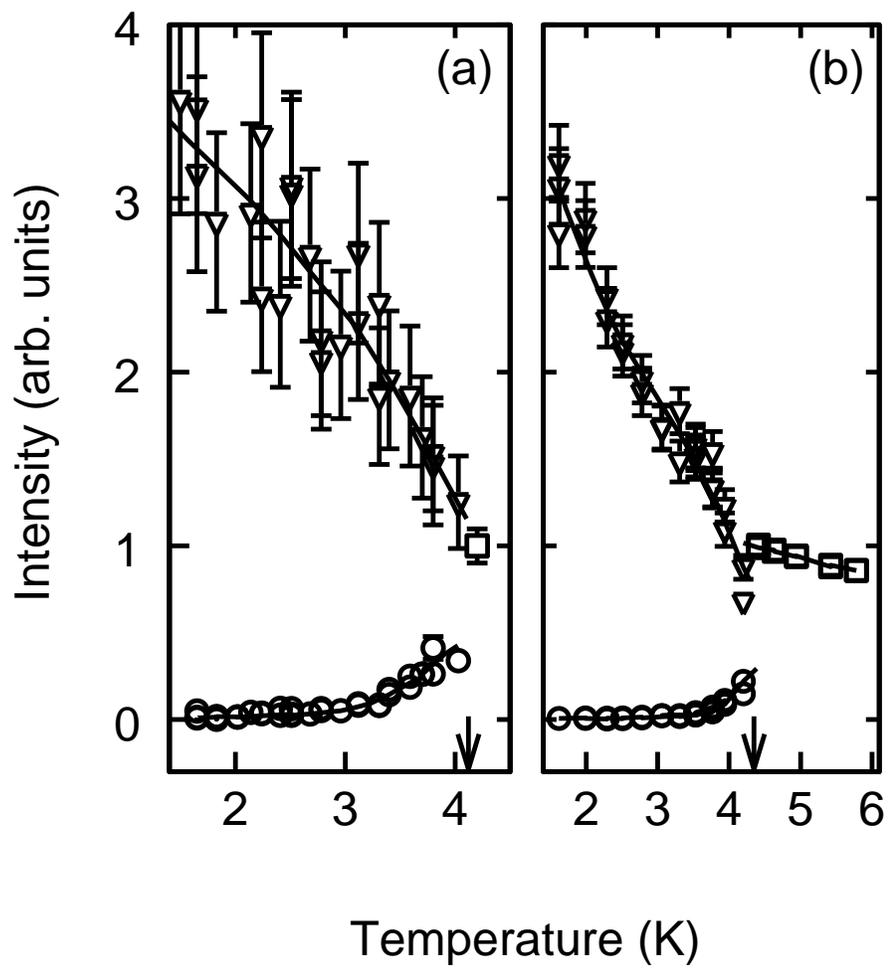
\item[Fig.8]   Temperature   dependences    of    the
intensities  of spectral components for  the  samples
with impurity concentrations 2.88\% (a), 3.20\% (b).
The    signs:    $\Box$-above   $T_N$,   $\bigtriangledown$
correspond   to   the   antiferromagnetic   resonance   and
$\bigcirc$ to the paramagnetic resonance below $T_N$. Solid
lines are guide-to-eyes. The N\'{e}el temperatures are marked
by arrows.

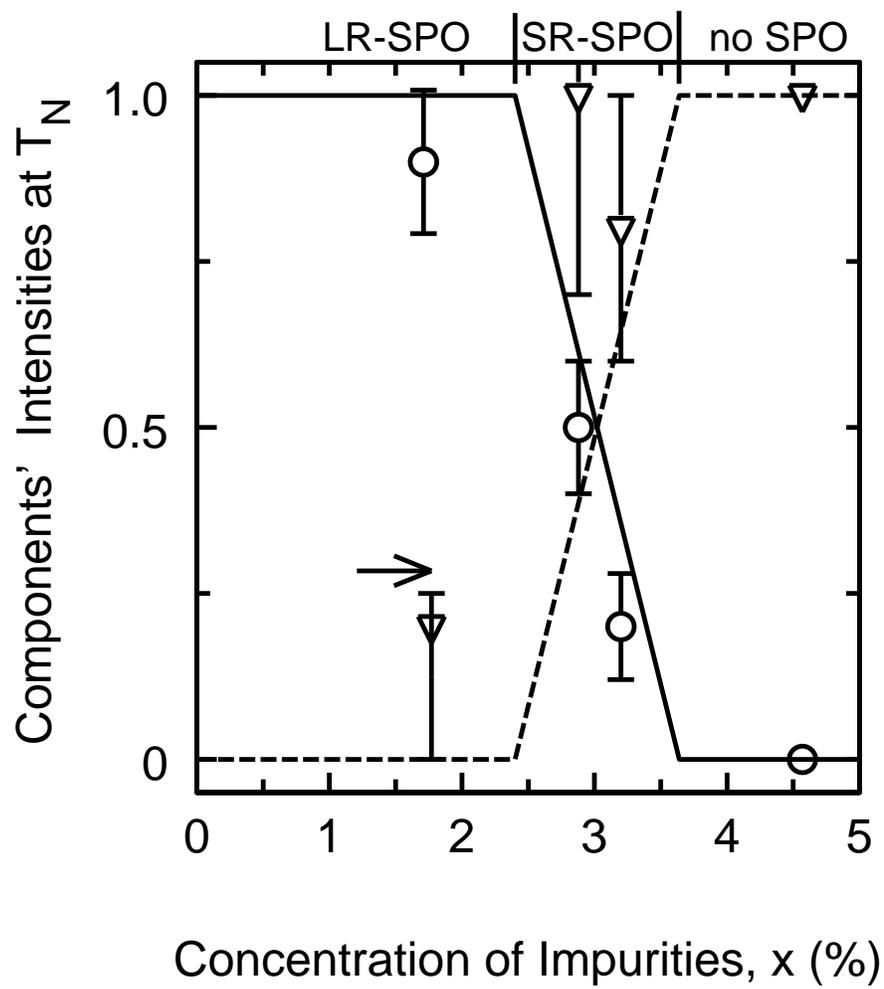
\item[Fig.9] Relative intensities of antiferromagnetic and
paramagnetic ESR components just below $T_N$ for different
concentrations $x$. The lines are linear interpolations between the
uniform and long range spin-Peierls ordering phases.
The arrow indicates the volume fraction of interpenetrating
spheres at the percolation point according to percolation theory.
\cite{ShanteKirkpatrick}

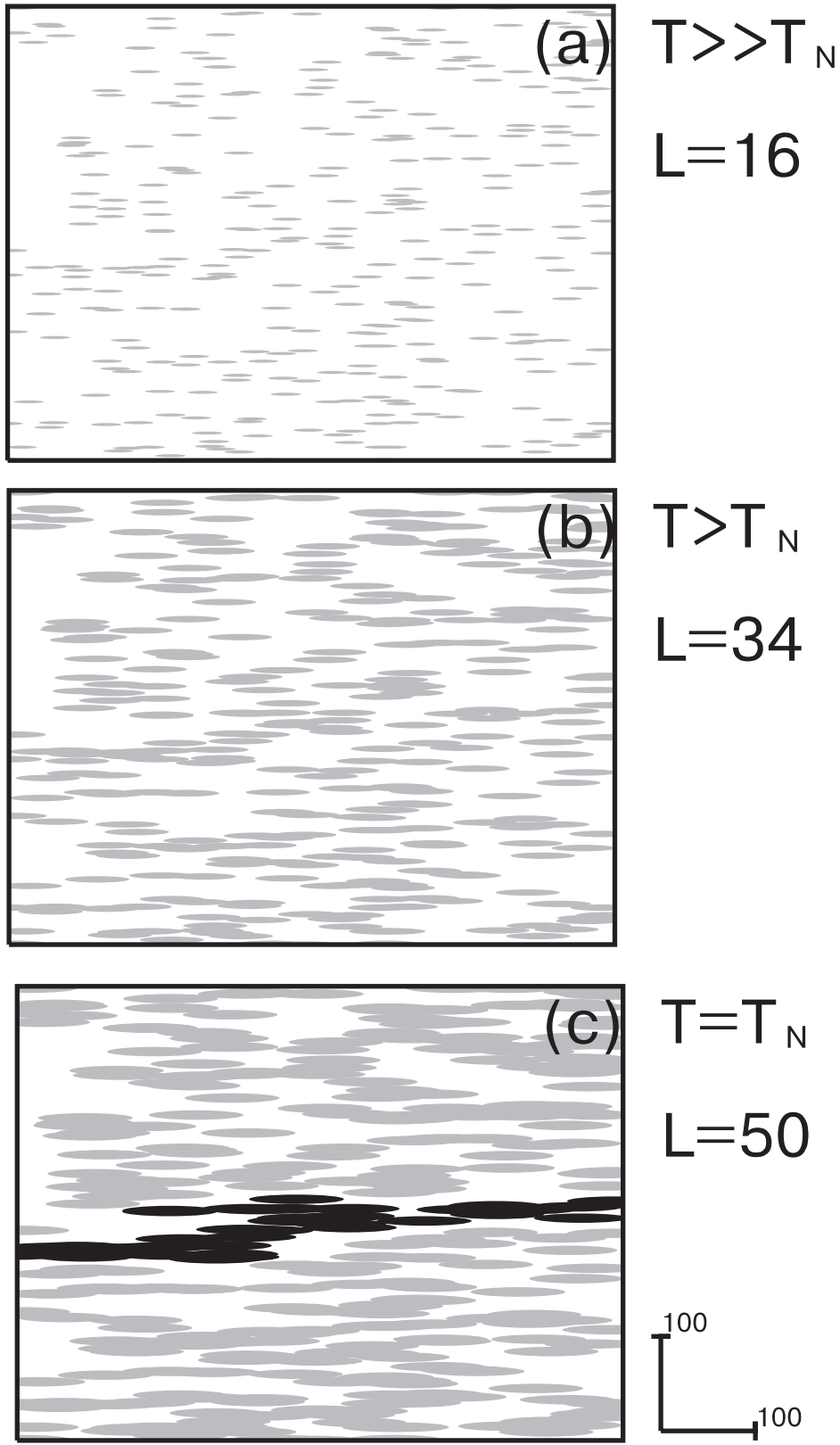
\item[Fig.10] Illustration of the
two dimensional modeling  of
the  formation  of the long-range antiferromagnetic  order.
Spin  chains  are  directed horizontally, drops  of  the
correlated  spins  are shown by grey filling,  spin-Peierls
matrix  is  white,  the macroscopic group  of  drops  is
marked  by  black filling.  The scale is given  in  interspin
distances.  The modeling is performed for $x=$0.1\% and the
following  values  of  $L$  (in  interspin distances):  {\it
a)}$L=$16, {\it b)} $L=$34; {\it c)}  $L=$50.

\end{itemize}

\end{document}